\pdfoutput=1
\documentclass
[twocolumn,aps,prd,amsmath,amssymb,floatfix]
{revtex4-1}
\usepackage{CJK}
\usepackage{graphicx}
\usepackage{bm}
\usepackage{mathptmx}
\usepackage{dcolumn}
\usepackage{multirow}
\usepackage{xcolor}
\usepackage[
breaklinks,pdfstartview=FitH,CJKbookmarks=true,
bookmarksnumbered=true,bookmarksopen=true,pdfborder={0 0 1},
colorlinks=true,linkcolor=blue,urlcolor=blue,anchorcolor=blue,citecolor=blue]
{hyperref}

\begin{document}

\def\be{\begin{equation}} \def\ee{\end{equation}}
\def\bal#1\eal{\begin{align}#1\end{align}}
\def\om{\omega}
\def\ms{M_\odot}
\def\mmax{M_\text{max}}
\def\ra{\rightarrow}
\def\de{\Delta}
\def\fm3{\;\text{fm}^{-3}}
\def\gc3{\,\text{g/cm}^3}
\def\rdu{\rho_\text{DU}} \def\xdu{x_\text{DU}} \def\mdu{M_\text{DU}}
\def\r1s0{\rho_{1S0}} \def\m1s0{M_{1S0}}

\title{
Neutron star cooling with medium-modified Urca rates
}

\begin{CJK*}{UTF8}{gbsn}

\author{B. X. Zhou (周博修)$^1$}
\author{J. B. Wei (魏金标)$^2$}
\author{Z. H. Li (李增花)$^{1,3}$}\email{zhli09@fudan.edu.cn}
\author{G. F. Burgio$^4$}
\author{H.-J. Schulze$^4$}

\affiliation{
$^1$Institute of Modern Physics,
Key Laboratory of Nuclear Physics and Ion-Beam Application, MOE,
Fudan University, Shanghai 200433, China
}
\affiliation{
$^2$Physics Department, University of Geosciences, Wuhan, P.R.~China
}
\affiliation{
$^3$Shanghai Research Center for Theoretical Nuclear Physics, NSFC,
Fudan University, Shanghai, 200438, China
}
\affiliation{
$^4$INFN Sezione di Catania, Dipartimento di Fisica,
Universit\'a di Catania, Via Santa Sofia 64, 95123 Catania, Italy}

\date{\today}

\begin{abstract}
We study the effects of in-medium modification of the elementary cooling
processes on observable properties of isolated neutron stars.
We deduce the neutron star mass distributions compatible
with the cooling analysis and compare with current theoretical models.
We conclude that current cooling data require fast direct Urca cooling,
moderated by proton superfluidity,
to be active in most neutron stars,
and that the DU onset threshold must lie below canonical masses.
In that case medium modifications of modified Urca rates
are practically insignificant,
but the $nn$ Bremsstrahlung rate plays a dominant role.
\end{abstract}

\maketitle
\end{CJK*}

\section{Introduction}

The cooling properties of neutron stars (NSs),
observationally accessible in terms of
temperature (or luminosity) vs age relations,
are an important tool to obtain a glimpse
on the internal structure and composition of NS matter
\cite{Yakovlev01,Yakovlev14,Miller21b,Burgio21}.
This information is complementary to global observables like
gravitational mass, radius, and tidal deformability,
accessible by other observational methods.

Together, these combined data of a single NS ideally
would be able to constrain the relevant equation of state (EOS) of NS matter.
Currently, due to the scarcity of such combined data,
and also the ambiguities and uncertainties of theoretical models for the EOS,
this goal has not been achieved.
However, recently great progress has been made
regarding observational information,
and cooling data are now available for about 60 objects
\cite{Potekhin20,Cooldat},
while global observables have been strongly constrained
by gravitational-wave observations in particular
\cite{Abbott18,Radice18,Raaijmakers21,Burgio21}.
This has allowed to further restrict the currently `valid'
theoretical EOSs \cite{Wei20c,Burgio21}.

This article is an attempt to update the cooling calculations
to the new data available,
both regarding cooling curves and theoretical EOS,
following our previous articles on this topic
\cite{Taranto16,Fortin18,Wei19,Wei20,Wei20b,Das24,Zhou25}.
In particular,
(a) it has become increasingly clear
\cite{Beznogov15,Beznogov15b,Potekhin20,Burgio21,Leinson22}
that fast neutrino cooling processes are required in order to explain
cold and not too old objects,
while also sufficiently slow cooling must be accommodated theoretically
to cover old and warm objects;
(b) the permissible nuclear EOS is more constrained now in terms of
maximum mass, radius, and tidal deformability,
so that several EOSs used in the past for cooling calculations
are not suitable any more.
We mention in particular the previously widely used APR EOS
\cite{Akmal98,Burgio18,Wei19b}.

Another particular feature of our work is the fact that now the
number of available cooling data is becoming sufficiently large
to allow a combined analysis of cooling properties
and NS mass distributions,
as initiated in \cite{Popov06,Beznogov15,Beznogov15b,Wei19,Wei20,Das24}.
This permits in particular to draw conclusions regarding the
superfluid properties of NS matter,
and is an important objective of the present article as well.

Finally, we now also investigate the effect of recently proposed
in-medium modifications of the modified Urca (MU) cooling rates
\cite{Voskresensky86,Blaschke95,Blaschke13,Niri16,Shternin18,
Suleiman23,Sedrakian24,Alford24}
on the cooling evolution and the combined data analysis.

This paper is organized as follows.
In Sec.~\ref{s:eos} we give a brief overview of the theoretical framework,
regarding the nuclear EOSs,
the various cooling processes,
and the related nucleonic pairing gaps.
Sec.~\ref{s:res} is devoted to the presentation and discussion
of the cooling diagrams and their dependence on the various
theoretical degrees of freedom.
In particular,
in Sec.~\ref{s:cor} we analyze the relation
between cooling diagrams and NS mass distributions,
quantifying the agreement of the deduced and theoretical mass distributions,
in order to constrain the pairing gaps in the stellar matter.
Sec.~\ref{s:med} investigates the impact of in-medium modifications.
Conclusions are drawn in Sec.~\ref{s:end}.
We use natural units $k_B=G=c=\hbar=1$.

\section{Formalism}
\label{s:eos}

In this work we employ a purely nucleonic scenario,
where NS matter is composed of nucleons and leptons only.
Exotic components like hyperons or quark matter are not considered
\cite{Burgio21}.
Even in this case,
the solution of the many-body problem for nuclear matter is still
a very challenging theoretical task.
A very rich literature does exist on this topic,
and the interested reader is referred to the recent
Refs.~\cite{Oertel17,Burgio21}
for details on the current state of the art.

\subsection{Cooling processes}

In this section we briefly recall the main neutrino emission mechanisms
in the NS core,
which dominate the total neutrino luminosity $L_\nu^\infty$.
In the context of NS cooling the one key property of the nuclear EOS
is whether it allows fast direct Urca (DU) cooling
by a large enough proton fraction
\cite{Yakovlev01,Page06,Page06b,Lattimer07,Yakovlev14,Potekhin15,Burgio21}.
The DU process is by several orders of magnitude
the most efficient one among all possible cooling reactions
involving nucleons and neutrino emission,
and depends on the NS EOS and composition.
The DU process involves a pair of charged weak-current reactions,
\be
 n \ra p + l + \bar{\nu}_l
\quad \text{and} \quad
 p + l \ra n + \nu_l \:,
\label{e:DU}
\ee
being $l=e,\mu$ a lepton and $\nu_l$ the corresponding neutrino.
Those reactions are allowed by energy and momentum conservation
\cite{Lattimer91}
only if $k_F^{(n)} < k_F^{(p)} + k_F^{(l)}$,
where $k_F^{(i)}$ is the Fermi momentum of the species $i$.
This implies that the proton fraction should be larger than a threshold value
$\xdu$ (about 13\% when including muons)
for the DU process to take place,
and therefore the NS central density should be larger
than the corresponding threshold density $\rdu$.

Thus different EOSs predict different DU threshold densities \cite{Burgio21}.
We choose here two microscopic Brueckner-Hartree-Fock (BHF) EOSs
obtained with the Argonne V18 \cite{Wiringa95}
or the Bonn~B (BOB) nucleon-nucleon ($NN$) potential \cite{Machleidt}
with compatible three-body forces
\cite{Li08a,Li08b,Liu22,Liu23},
see \cite{Baldo99,Baldo12} for a more detailed account.
These EOSs are compatible with all current low-density constraints
\cite{Wei20b,Burgio21,Burgio21b}
and in particular also with those imposed on NS maximum mass
$\mmax>2\ms$
\cite{Antoniadis13,Arzoumanian18,Cromartie20},
radius $R_{1.4}\approx11$--$13\,$km
\cite{Riley21,Miller21,Pang21,Raaijmakers21,Rutherford24,Mauviard25},
and tidal deformability $\Lambda_{1.4}\approx70$--$580$
\cite{Abbott17,Abbott18,Burgio18,Wei19}.

The BHF method provides the EOS for homogeneous nuclear matter,
and therefore an EOS for the low-density inhomogeneous crustal part
has to be added.
For that,
we adopt the well-known Negele-Vautherin EOS \cite{Negele73}
for the inner crust in the medium-density regime
($0.001\fm3 < \rho < \rho_t$),
and the ones by Baym-Pethick-Sutherland \cite{Baym71} and
Feynman-Metropolis-Teller \cite{Feynman49} for the outer crust
($\rho < 0.001\fm3$).
By imposing a smooth transition
of pressure and energy density between both branches of the betastable EOS
\cite{Burgio10},
one finds a transition density at about $\rho_t \approx 0.08\fm3$.
In any case the NS maximum mass domain is not affected by the crustal EOS,
with a limited influence on the radius and related deformability
for NSs with canonical mass value
\cite{Burgio10,Baldo14,Fortin16,Tsang19}.

If the DU process is kinematically forbidden or strongly reduced,
various much less efficient neutrino processes may be operating in the NS core.
While the former is a pure weak reaction,
those processes involve nucleon collisions driven by strong interactions,
and are consequently affected by much greater theoretical uncertainties
that will be discussed in the following.
The two main ones with an emissivity $Q_\nu\propto T^8$ are the MU processes,
\be
 n + N \ra p + N + l \! + \bar{\nu}_l
\quad \text{and} \quad
 p + N + l \ra n + N + \nu_l \:,
\label{e:MU}
\ee
where $N=n,p$ is a spectator nucleon that ensures momentum conservation.
Since five degenerate fermions are involved instead of three,
the efficiency is significantly reduced compared to the DU process.

The $NN$ bremsstrahlung (BS) reactions,
\be
 N + N \ra N + N + \nu+\bar{\nu} \:,
\ee
with $N$ a nucleon and $\nu$, $\bar{\nu}$ an (anti)neutrino of any flavor,
are also abundant in NS cores,
and their rate increases with the baryon density,
but they are orders of magnitude less powerful than the DU or MU one,
thus producing a slow cooling \cite{Yakovlev01}.
All those cooling mechanisms can be strongly affected
by the superfluid properties of the stellar matter, i.e.,
critical temperatures and gaps in the different pairing channels.
We will turn to this theoretical issue in Sec.~\ref{s:gap}.

\subsection{Modified Urca cooling}

In this section we provide some details on MU cooling,
in particular the relevance of the nucleon effective masses
and other in-medium effects,
following closely the detailed treatment given in \cite{Yakovlev01}
and updates in \cite{Potekhin15,Schmitt18,Bottaro24,Kopp24}.
Only the rates for the non-superfluid scenarios will be given in this section,
for which the dependence on the effective masses is via the general factor
\cite{Baldo14b}
\be
 M_{ij} \equiv
 \bigg(\frac{m_n^*}{m_n}\bigg)^i \bigg(\frac{m_p^*}{m_p}\bigg)^j
 \bigg( \frac{\rho_p}{\rho_0} \bigg)^{\!\!1/3}
 \frac{k_l}{\mu_l}  \:.
\label{e:m}
\ee
In the presence of superfluidity the dependence becomes highly nontrivial
and requires detailed calculations \cite{Yakovlev01}.

The emissivities of the MU processes in the neutron and proton branches,
employing for the long-range part of the in-medium $NN$ interaction
the dominant one-pion-exchange (OPE) contribution and
for the short-range part an effective contribution
in the framework of Landau theory,
are given by  
\cite{Friman79,Yakovlev95,Yakovlev01,Potekhin15,Bottaro24,Kopp24}
\bal
 Q^{(Mn)} \approx\;& 8.55\times10^{21} M_{31} T_9^8 \alpha_n \beta_n \:,
\label{e:mun}
\\
 Q^{(Mp)} \approx\;& 8.55\times10^{21} M_{13} T_9^8 \alpha_p \beta_p
\nonumber\\&\times
 \frac{(k_l + 3k_p - k_n)^2}{8 k_l k_p}
 \mathop\Theta( k_l + 3k_p - k_n ) \:,
\label{e:mup}
\eal
in units of erg$\,$cm$^{-3}$s$^{-1}$,
where $k_i \equiv k_F^{(i)}$, $i=e,\mu,n,p$,
and the factors $\alpha_n,\alpha_p$
take into account the momentum-transfer dependence
of the squared reaction matrix element under the Born approximation,
and $\beta_n,\beta_p$ include the non-Born corrections
due to $NN$ interaction effects,
which are not described by the OPE contribution \cite{Yakovlev01}.
The currently adopted values are
$\alpha_p=\alpha_n=1.13$ in the original Friman-Maxwell (FM) calculation
\cite{Friman79},
whereas the more accurate momentum-transfer average discussed in
\cite{Yakovlev01,Yakovlev95} provides
\be
 \alpha_p=\alpha_n=1.76-0.63(\rho_0/\rho_n)^{2/3} \:.
\label{e:mua}
\ee
This is the expression implemented in the cooling code
and used throughout this work, with
$\beta_p=\beta_n=0.68$ in both cases.

The main difference between the proton branch and the neutron branch
is the threshold character,
although usually irrelevant.
If muons are present in the dense NS matter,
the equivalent MU processes become also possible,
as discussed in Refs.~\cite{Yakovlev01,Kopp24,Bottaro24}.

The problem of an accurate computation of the MU rates
can currently be considered unresolved,
as no formalism incorporating consistently
(in particular, compatible with the computation of the EOS)
the various in-medium effects has yet been attempted.
It seems, however, that the simple expressions Eqs.~(\ref{e:mun},\ref{e:mup})
constitute underestimates of the real MU rates.
We discuss now briefly the current status of such attempts.

\subsection{Medium modifications}

The standard cooling rates mentioned above are computed by assuming
reaction matrix elements in the quasiparticle approximation
(with sharp Fermi surfaces, apart from temperature effects)
at the lowest order in density and temperature,
and employing a simple approximation
(free OPE + Landau parameters)
for the in-medium $NN$ interaction.
They are consequently density independent
apart from the effective-mass corrections and kinematic factors,
but are subject to potentially very important in-medium corrections.
This is a very complex topic that has been extensively studied in the past,
without coming to definite quantitative conclusions so far.

The motivation of our work is to include in the cooling calculations
some of the most recent theoretical estimates of medium effects,
in order to probe the sensitivity of the final results to them
and to identify observable phenomena that might be able to constrain those.

First, let us note that a recent detailed reevaluation of the FM results
\cite{Friman79}
led to the claim that those are too large by a factor about 2
\cite{Bottaro24}.

While earlier works
\cite{Voskresensky86,Blaschke95,Blaschke13,Niri16}
tried to improve the modeling of the
$NN$ (OPE) interaction used in the FM results,
coming to quite different conclusions
regarding enhancement or suppression of the MU rate,
recent theoretical studies focused on the importance of the nucleon
in-medium propagation.

Several works attempted to go beyond the approximation of degenerate Fermi
gases with sharp momentum distributions
by including empirically the quasi-particle $Z$-factor \cite{Dong16}
or extended tails of the momentum distributions \cite{Frankfurt08}.
However, consistent modifications of the effective masses
or the $NN$ interaction were disregarded.

\cite{Shternin18,Suleiman23}
investigated the importance of the nucleon off-shell propagator,
which is approximated by a constant in the standard MU treatment
leading to Eqs.~(\ref{e:mun},\ref{e:mup}),
thus neglecting the appearance of a pole above the DU threshold.
They concluded that this generic kinematic effect
can enhance the MU rate by a large density-dependent factor,
in particular close to the DU onset, 
and thus provides a more smooth switch-on of the DU process.
On the other hand,
their use of a $G$-matrix to replace the simple $NN$ interaction
employed by FM resulted in a substantial decrease of the rate,
in qualitative agreement with the in-medium $T$-matrix result of
\cite{Blaschke95}.

The overall result of \cite{Shternin18}
for the MU enhancement factor $R(\rho)$
was parameterized in the following convenient form:
\bal
 R &= 0.6 R_A + 0.2 R_B \:,
\label{e:sbh}
\\
 R_A &= \frac{2{m_p^*}^2{\mu_l}^2}{k_l}
 \int_{k_n-k_l}^{k_n+k_l} \frac{dk}{(k_p^2-k^2)^2} \:,
\\
 R_B &= \frac{2{m_n^*}^2{\mu_l}^2}{k_l} \int_{k_p-k_l}^{k_p+k_l}
 \frac{dk\; k/k_p}{(k_n^2-k^2)^2} \:,
\eal
with $l=e,\mu$,
which only requires the knowledge of the density dependence of the various
Fermi momenta $k_i \equiv k_F^{(i)}$ for a given EOS.
Here $R$ is normalized to the free-OPE rate,
so that $R=1$ corresponds to the free-OPE matrix element,
which in the standard cooling code
is corrected by the factor $\alpha_n\beta_n$, Eq.~(\ref{e:mun}).
The enhancement factor remains valid when including
standard superfluidity suppression factors for the MU rate.

\cite{Sedrakian24} included the width of the proton spectral function
due to finite temperature and short-range nuclear correlations
into the computation of the DU rate,
which leads to a smoothening of the sharp onset of DU cooling
with increasing proton fraction.
The DU rate becomes then comparable to the MU rate
at densities below the standard DU onset.
However, the effect of pairing has so far not been included consistently
in this scenario.

\cite{Alford24} tried to improve the MU rate by including the width of
the nucleon propagator via its imaginary part.
This leads to a combined treatment of DU and MU processes,
in which the MU rate is substantially enhanced compared
to the standard expression.
The enhancement is smaller than that obtained in \cite{Shternin18},
and only at very low density both methods agree.
However, the procedure is not compatible with that of \cite{Sedrakian24}
and the numerical predictions are quite different.
Also in this case, superfluidity was disregarded.

Thus,
so far no consistent combined treatment of DU and MU rates has been developed.
Different works have focused on improving particular specific features
of the standard results,
without establishing fully consistent procedures,
in particular regarding the in-medium $NN$ interaction.
In this situation,
we employ in this work the parametrization of Eq.~(\ref{e:sbh}) in order
to estimate upper limits of the MU rate enhancement.
Of course such enhancement will only have visible effects in NSs
without active DU cooling,
which is still several orders of magnitude stronger
than any enhanced MU cooling.
This feature and in particular the NS mass range without active DU cooling
depends on the nuclear EOS that is discussed in the next subsection.

\begin{figure}[t]
\vspace{-4mm}
\centerline{\includegraphics[scale=0.61]{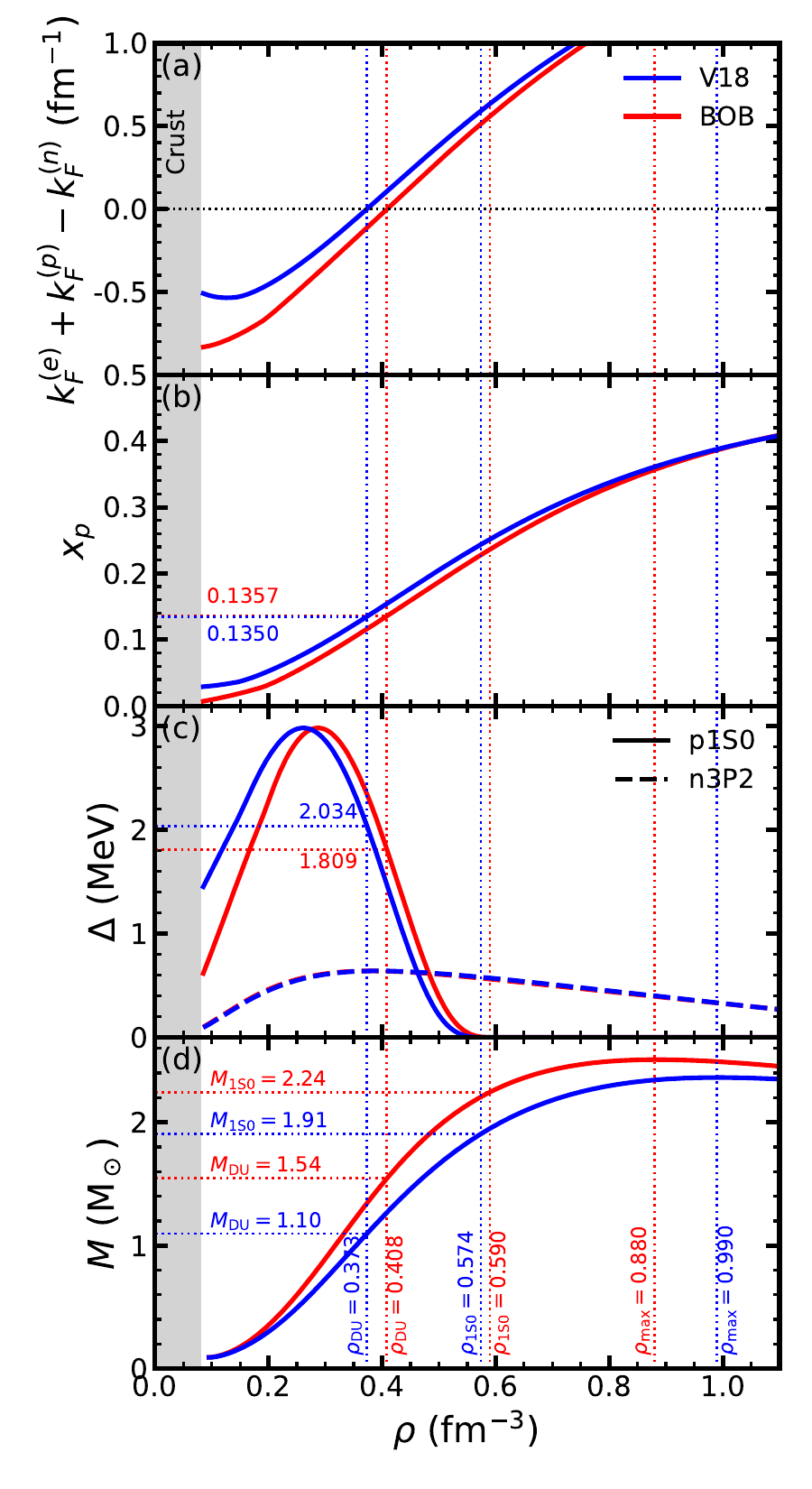}}
\vspace{-7mm}
\caption{
The threshold condition for the DU$e$ process (a),
the proton fraction (b),
the p1S0 and n3P2 BCS gaps (c),
and the gravitational mass (d)
vs.~the nucleon (central) density for V18 and BOB EOSs.
Dotted lines
indicate the onset of DU cooling $\rdu$,
the vanishing of the p1S0 gap $\r1s0$, and
the $\mmax$ configurations.
}
\label{f:xp}
\end{figure}

\subsection{Equation of state}
\label{s:du}

In order to illustrate the relevant properties of the chosen EOSs,
we show in Fig.~\ref{f:xp} the electron DU onset condition (a)
and the proton fraction (b),
which reaches the DU threshold $\xdu \approx 0.135$
at a density $\rdu \approx 0.37(0.41)\fm3$
for the V18 (BOB) EOS.
The associated NS mass is $\mdu=1.10(1.54)\ms$,
hence all heavier NSs can potentially cool very fast.

The DU thresholds for muons lie slightly higher,
$\rho_{\text{DU}\mu} \approx 1.13(1.11)\,\rdu$
and $M_{\text{DU}\mu} \approx 1.21(1.15)\,\mdu$
for the V18 (BOB) EOS,
see also Fig.~\ref{f:mmu}.
This difference is relevant for the MU cooling,
but plays an important role only for the BOB EOS in the range $M<1.54\ms$,
whereas it is practically irrelevant for the V18 EOS,
where DU cooling always dominates.
The consequences will be analyzed in the following.

We stress that the values of the maximum mass $\mmax=2.36(2.51)\ms$,
Fig.~\ref{f:xp}(d),
are compatible with the current observational lower limits
\cite{Antoniadis13,Arzoumanian18,Fonseca21},
in particular
the recent data $M=2.35\pm0.11\ms$ for PSR J0952-0607 \cite{Romani25}.
However, only the V18 EOS
\cite{Sun21,Burgio21,Wei20}
($R_{1.4},R_{2.0},R_{2.08} = 12.3,12.0,11.9\,$km)
is also fully compatible with the limits on the radii
$R_{2.08}$ \cite{Miller21},
$R_{2.0}$ \cite{Rutherford24,Mauviard25}, and
$R_{1.4}$
\cite{Pang21,Raaijmakers21,Rutherford24,Dittmann24,Mauviard25,Miller26},
obtained from combined NICER analyses of the pulsars
J0030+0451, J0740+6620, J0437-4715, J0614-3329, and J1231-1411,
together with GW170817,
whereas the radii obtained with the BOB EOS
($R_{1.4},R_{2.0},R_{2.08} = 12.9,12.8,12.7\,$km)
are slightly too large.
We therefore consider the V18 EOS more realistic for several reasons,
which will also become clear in the following.
We use the BOB EOS here only in order to demonstrate the effects of a
large DU onset density and mass.

\subsection{Pairing gaps}
\label{s:gap}

One of the most important nuclear physics input for the NS cooling simulations
are the superfluid properties of stellar matter,
basically the neutron and proton pairing gaps
in the different reaction channels
\cite{Yakovlev01,Sedrakian19,Burgio21}.
The most important ones are the proton 1S0 (p1S0)
and neutron 3PF2 (n3P2) pairing channels;
the proton 3PF2 gap is often neglected due to its uncertain properties
at large densities,
while the neutron 1S0 gap in the crust is of little relevance for the cooling.
Moreover, in our previous works we found that nonzero values of the n3P2 gap
cannot reproduce the current observational data in our cooling simulations
\cite{Wei20,Wei20b,Das24},
so we exclude it in this work
and focus our study on the possible p1S0 gap function $\Delta(\rho)$.

We use the same method as in our previous works
\cite{Taranto16,Wei20,Wei20b,Das24,Zhou25},
namely a rescaling of both magnitude and range of the naive BCS gap
in terms of global scaling factors $s_y$ and~$s_x$,
\be
 \Delta(\rho) \equiv s_y \Delta_\text{BCS}(\rho/s_x) \:,
\label{e:sxy}
\ee
which will be considered as free parameters in the cooling calculations.
Their optimal values will be determined later
in a combined analysis of cooling data and deduced NS mass distributions.

We remind that superfluidity arises from the formation of
$pp$ and $nn$ Cooper pairs mediated by the attractive component
of the $NN$ interaction.
The onset of pairing is characterized by a critical temperature $T_c$,
related to the pairing gap $\Delta$ through $T_c \simeq 0.567 \Delta$.
For temperatures below $T_c$,
the presence of pairing has two main consequences for neutrino emission.
First, neutrino-emission processes involving the paired component
are exponentially suppressed.
Second, the formation and breaking of Cooper pairs
gives rise to the pair-breaking and formation (PBF) process,
accompanied by the emission of neutrino-antineutrino pairs.
The PBF emissivity becomes non-zero when the temperature
drops below the critical temperature of the corresponding baryon species,
reaches its maximum efficiency at $T\simeq 0.8 T_c$,
and is exponentially suppressed at temperatures $T\ll T_c$.

Further important ingredients in the cooling simulations
are the neutron and proton effective masses,
which we actually used in \cite{Taranto16}.
In the BHF approach,
the effective masses can be computed consistently
together with the EOS
in terms of the s.p.~energy $e(k)$ \cite{Baldo14b},
\be
 \frac{m^*(k)}{m} = \frac{k}{m} \left[ \frac{d e(k)}{dk} \right]^{-1} \:,
\ee
and convenient parametrizations can be found in \cite{Baldo14,Shang20}.

The active range of unscaled BCS p1S0 pairing is shown in Fig.~\ref{f:xp}(c).
For the V18 (BOB) EOS the gap vanishes at $\r1s0=0.57(0.59)\fm3$,
corresponding to the central density of a $\m1s0=1.91(2.24)\ms$ NS.
Therefore $M<\m1s0$ stars are superfluid throughout,
whereas for $M>\m1s0$ there is a growing non-superfluid core region
that is cooling very fast by the unquenched DU process.
However, the parameter $s_x$ changes the active range of pairing,
and is thus essential for the confrontation with cooling data
in the next section.

For completeness, we also display the unscaled n3P2 BCS gap
in Fig.~\ref{f:xp}(c).
It extends over the entire density and mass range,
and therefore would block all cooling processes involving neutrons for all NSs.
However, the competing n3P2 PBF process provides too strong cooling
for old objects,
in disagreement with some data,
as found in \cite{Grigorian05,Wei20,Das24}.

\section{Cooling simulations}
\label{s:res}

The NS cooling simulations are performed with the widely used
one-dimensional code {NSCool} \cite{Pageweb},
which relies on the implicit scheme of \cite{Henyey64}
to solve the general-relativistic equations of
energy balance and energy transport,
\bal
 & \frac{e^{-\lambda-2\phi}}{4\pi r^2}
 \frac{\partial}{\partial r}(e^{2\phi}L) + Q_\nu =
 - \frac{c_V}{e^\phi} \frac{\partial T}{\partial t} \:,
\\
 & \frac{L}{4\pi r^2} =
 -\kappa e^{-\lambda-\phi}
 \frac{\partial}{\partial r}(e^\phi T) \:.
\eal
Here, the local non-neutrino luminosity $L$, 
the neutrino emissivity $Q_\nu$, and
the temperature $T$
are functions of the radial coordinate $r$ and time $t$.
The metric functions are $\phi$ and
$e^{-\lambda}=\sqrt{1-2m/r}$
with the enclosed mass $m(r)$.
The specific heat capacity $c_V$ and the thermal conductivity $\kappa$ are
key microphysical inputs.
The observable quantities are the redshifted
surface temperature $T_s^\infty=T(R) e^{\phi(R)}$
and the photon luminosity at infinity
$L_\gamma^\infty=L_\gamma(R) e^{2\phi(R)}$.

The partial differential equations are solved on a grid of spherical shells.
For this purpose,
the star is divided into two regions separated by an outer boundary
located at radius $r_b$,
corresponding to a density $\rho_b=10^{10}\gc3$.
In the low-density region ($\rho<\rho_b$),
the so-called envelope \cite{Beznogov21},
which accounts for variations in mass and composition (e.g., due to accretion),
is treated separately within the code.
We adopt the envelope model of \cite{Potekhin97}.
In the region $\rho>\rho_b$ ($r<r_b$),
the matter is strongly degenerate,
and the structure of the star is therefore assumed to remain unchanged in time.
For a NS of given mass and envelope composition,
the simulation yields a set of cooling curves, i.e.,
the luminosity $L_\gamma^\infty$ as a function of the NS age $t$.
Each simulation is initialized with a uniform temperature profile,
$\tilde{T}\equiv Te^{\phi}=10^{10}\,$K,
and is terminated when $\tilde{T}$ falls to $10^4\,$K.

Neutrino emission is the dominant ingredient of the cooling process,
and the code incorporates all relevant reactions:
the nucleonic DU and MU processes,
nucleon-nucleon BS and PBF process,
including the modifications induced by p1S0 and n3P2 pairing.
The PBF rates have been updated according to \cite{Leinson10}.
In addition, several crustal processes are taken into account,
most notably electron-nucleus bremsstrahlung, plasmon decay,
electron-ion bremsstrahlung, etc.

\begin{figure*}[t]
\vspace{-4mm}
\centerline{\includegraphics[width=0.84\textwidth]{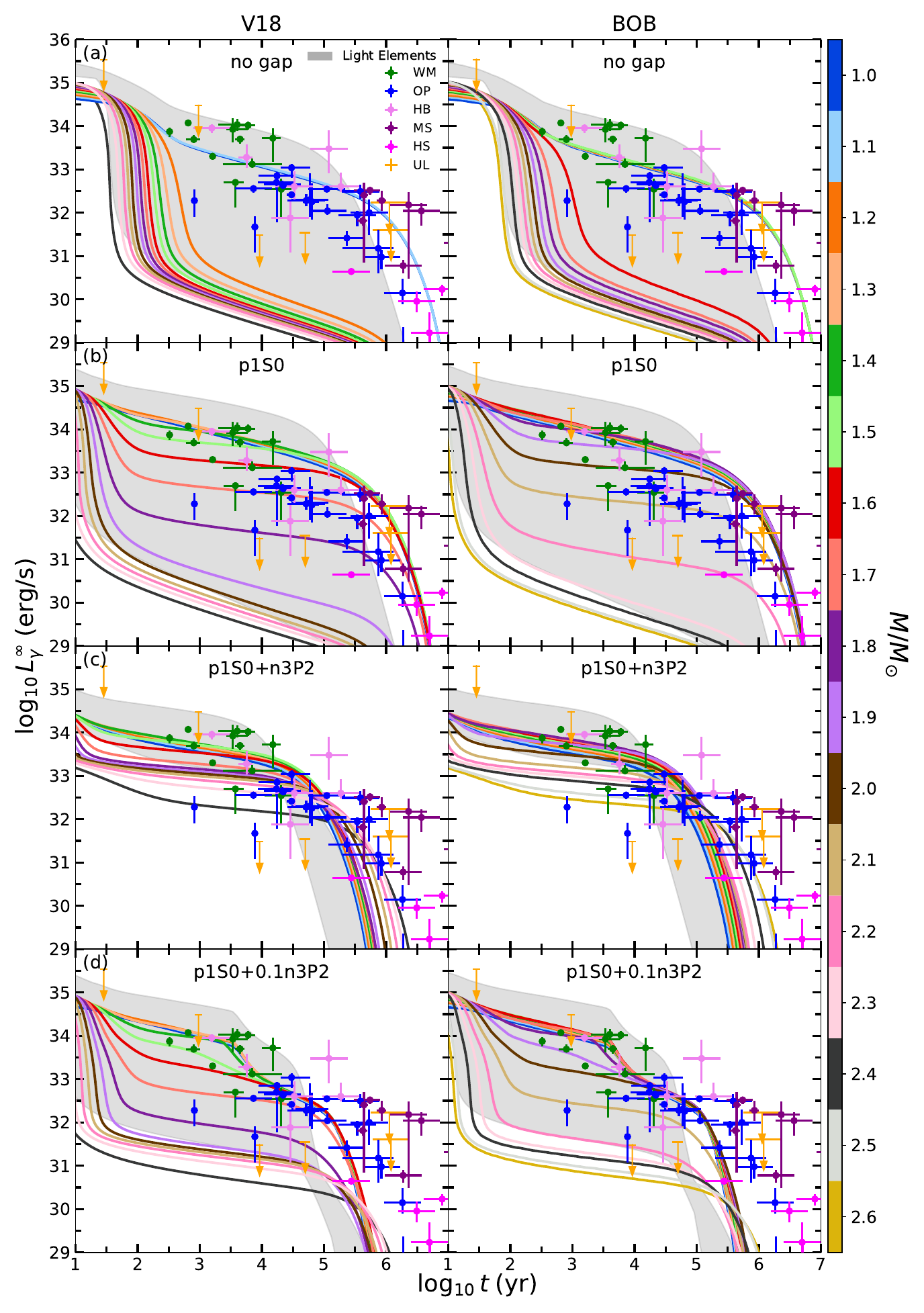}}
\vspace{-4mm}
\caption{
Cooling diagrams
obtained with V18 (left column) and BOB (right column) EOSs
(a) without superfluidity gaps (1st~row),
(b) with only p1S0 BCS gap (2nd row),
(c) with p1S0+n3P2 BCS gaps (3rd row), and
(d) with rescaled ($s_y=0.1$) n3P2 BCS gaps (lower row),
for different NS masses $M/\!\ms=1.0,1.1,\ldots,\mmax$
(decreasing curves).
The fixed-$M$ curves are obtained with a Fe atmosphere,
while the gray-shaded bands indicate
the domains covered by the corresponding light-elements (accreted)
atmosphere models.
The data points are from \protect\cite{Potekhin20,Cooldat}
and are grouped as
weakly-magnetized NSs (WM),
ordinary pulsars (OP),
high-magnetic-field pulsars (HB),
the magnificent seven (MS),
small hot spots (HS),
and upper limits (UL).
See text for more details.
}
\label{f:cool}
\end{figure*}

\begin{figure*}[t]
\centerline{\hspace{-0mm}\includegraphics[width=1.\textwidth]{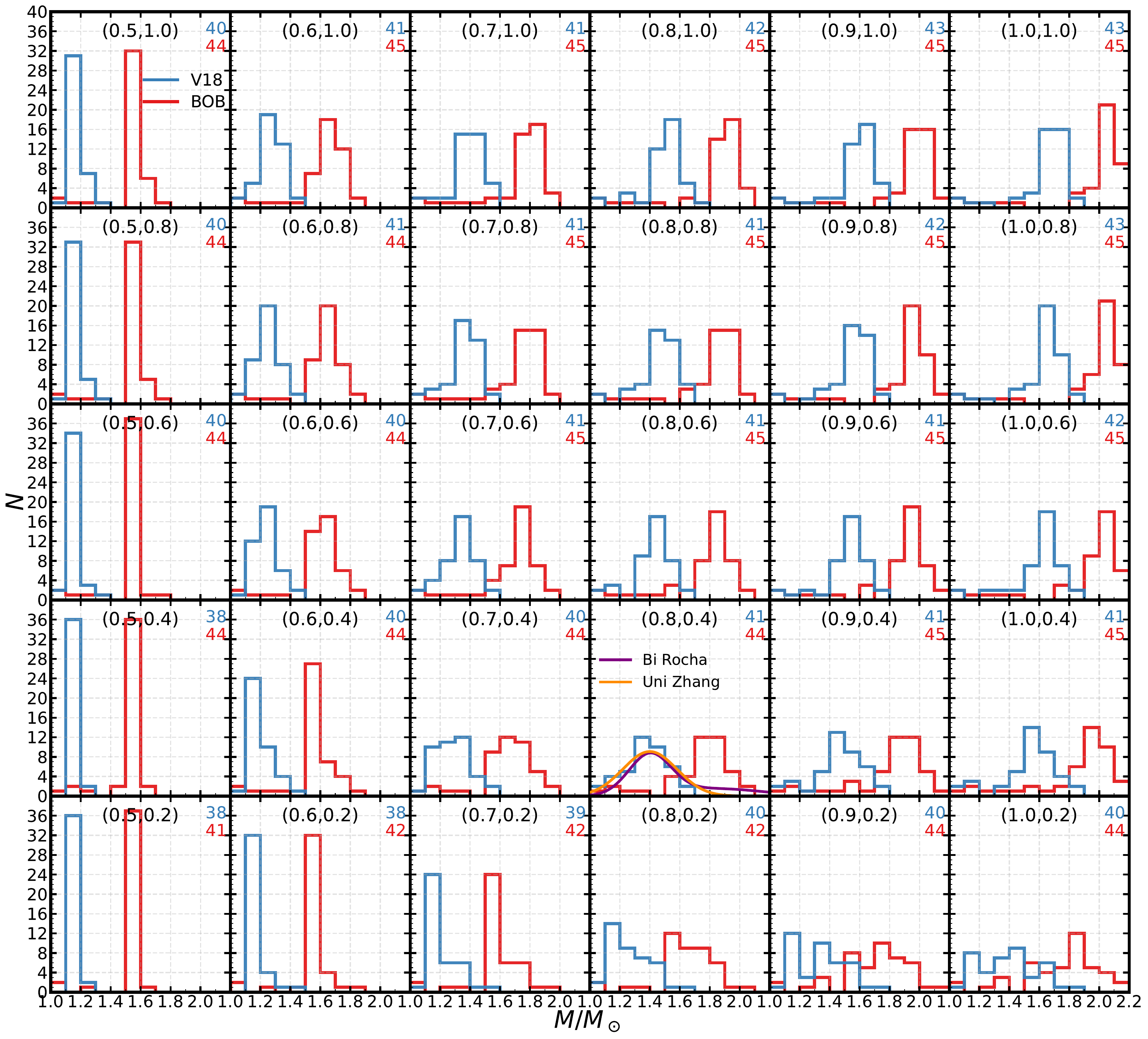}}
\vspace{-3mm}
\caption{
Deduced NS mass distributions for scaling factors
$(s_x=0.5,\ldots,1.0) \otimes (s_y=0.2,\ldots,1.0)$
and with Fe atmosphere,
for the V18 (blue) or BOB (red) EOS.
$N$ is the number of data points lying in a given mass interval
$\Delta M=0.1\ms$
in the proper ($s_x,s_y$) cooling diagram.
The (0.8,0.4) panel shows the best-fit theoretical results of
\cite{Zhang11,Rocha19}
for the V18 EOS superimposed.
The top-right numbers indicate the total number of data points
in the histograms.
}
\label{f:NSdist}
\end{figure*}

\begin{figure}[!t]
\vspace{-6mm}
\centerline{\includegraphics[width=0.52\textwidth]{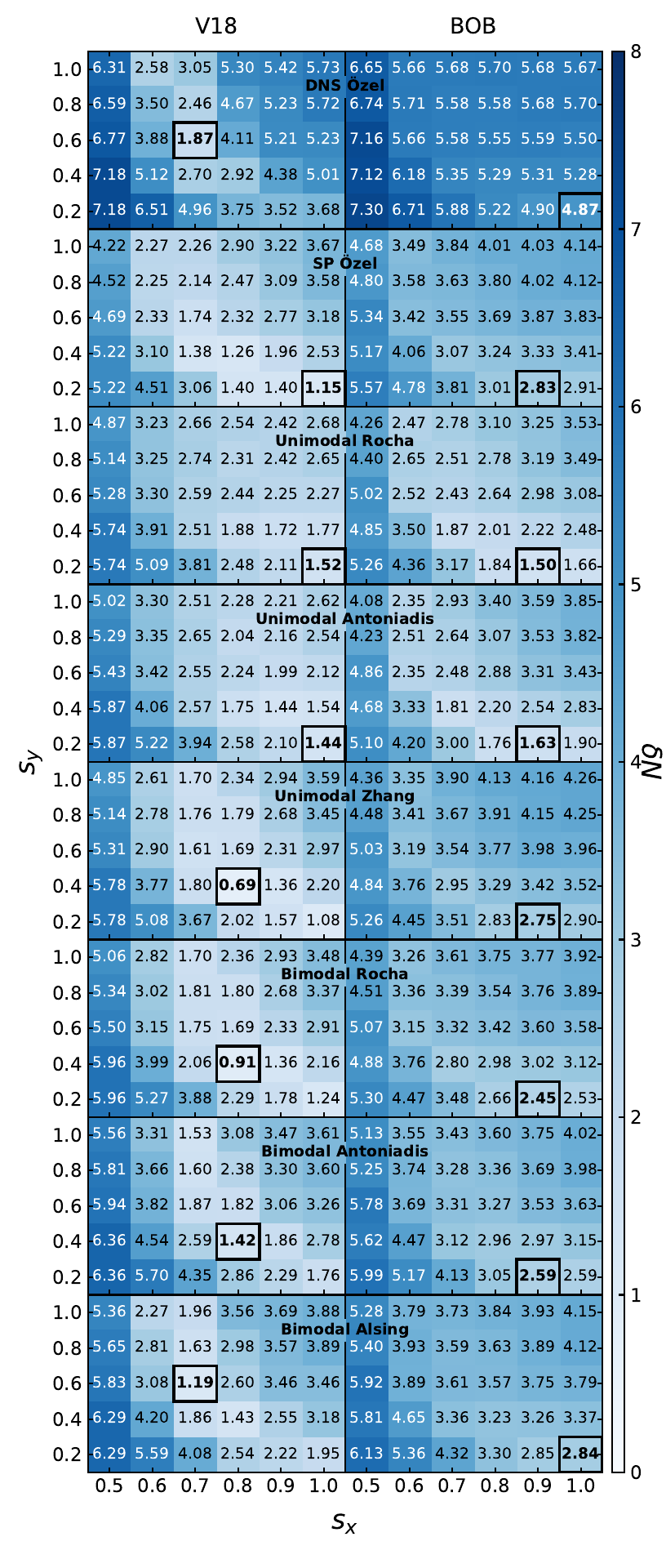}}
\vspace{-7mm}
\caption{
Root-mean-square variance $\delta N$, Eq.~(\ref{e:rms}),
between deduced NS mass distributions
with V18 or BOB EOS obtained with different p1S0 gap scale factors $s_x,s_y$
for a Fe atmosphere,
and various theoretical distributions
\cite{Zhang11,Ozel12,Kiziltan13,Antoniadis16,Alsing18,Rocha19,Landry21}.
Preferred values are indicated in squared boxes.
}
\label{f:heatmap}
\end{figure}

\subsection{Cooling curves}

Before examining the effect of gap scaling factors and in-medium corrections
we present results of the `standard' calculations
without such modifications.

In Fig.~\ref{f:cool} we show the $L_\gamma^\infty(t)$ cooling diagrams
obtained with a Fe atmosphere (individual curves for fixed NS masses)
or a light-elements atmosphere (gray-shaded bands),
in comparison with the currently known 59 observational data
\cite{Beznogov15,Potekhin20,Cooldat}.
Panels (a) display the results obtained without superfluidity.
One notes the dominant effect of the DU process,
characterized by a far too rapid cooling for all NSs with
$M>\mdu=1.10(1.54)\ms$ for V18 (BOB) EOS,
which is clearly unrealistic, regardless of the atmosphere model.
In panels (b) we include the p1S0 BCS ($s_x=s_y=1$) gap.
The main effect is the quenching of the DU process
such that stars in the overlap zone
$1.10(1.54)\ms = \mdu < M < \m1s0 = 1.91(2.24)\ms$
cool moderately fast now,
and only high-mass stars, $M>1.91(2.24)\ms$, cool very rapidly.
Now all data points can be covered within their error bars,
taking advantage of the two atmosphere models.
While the parameter $\mdu$ is fixed by the EOS model,
the value of $\m1s0$ depends on the scaling parameter $s_x$,
reflecting our current lack of knowledge regarding the p1S0 gap in dense matter.
This allows to tune the assignment of NS masses to the data points
in the quenched DU region,
which will be exploited in the following.

In panels (c) we investigate the effect of the n3P2 BCS gap:
This extends over the entire density range,
and therefore blocks the DU process for all NSs.
On the other hand the competing n3P2 PBF process provides a too strong cooling
for the old ($\gtrsim10^6\,$yr) objects \cite{Wei20b,Das24},
with or without DU process,
as already found by other authors
\cite{Grigorian05,Beznogov15,Beznogov15b,Beznogov18,Potekhin19,Potekhin20,Wei20}.
The discrepancy is even worse
when using a n3P2 BCS gap scaled down by a factor $s_y=0.1$
(i.e., with a maximum gap of about 0.06~MeV
\cite{Page11,Zhao25,Potekhin26}),
shown in panels (d).

In conclusion,
an EOS featuring DU cooling for a wide enough mass range of NSs
together with (partial) quenching by the p1S0 BCS gap
seems to be required to reproduce the cooling data.
It seems difficult to accommodate finite n3P2 pairing in this setup.
We therefore continue the analysis including p1S0 pairing
but without n3P2 pairing.

\begin{figure}[t]
\vspace{-3mm}
\centerline{\includegraphics[scale=0.65]{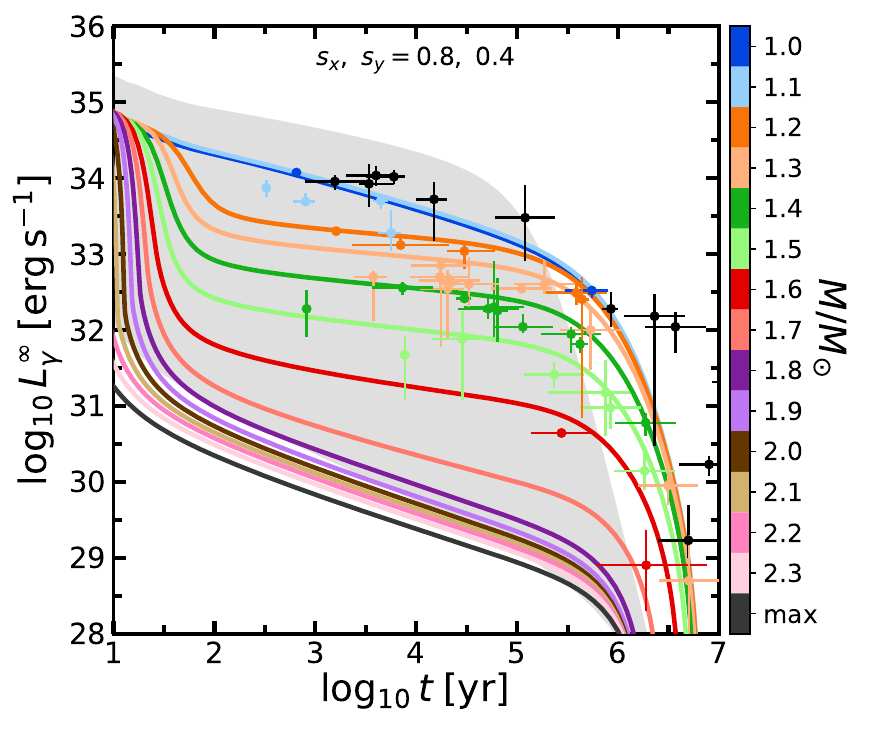}}
\vspace{-4mm}
\caption{
Cooling diagram
with the V18 EOS
for $(s_x,s_y)=(0.8,0.4)$.
The fixed-$M$ curves are obtained with a Fe atmosphere,
while the gray-shaded band indicates the domain covered by the
light-elements (accreted) atmosphere model.
Colors of the data points indicate their mass assignment,
corresponding to the closest upper fixed-$M$ cooling curve.
Black data are not in the mass histogram.
}
\label{f:best}
\end{figure}

\begin{figure*}[t]
\vspace{-4mm}
\centerline{
\includegraphics[scale=0.63]{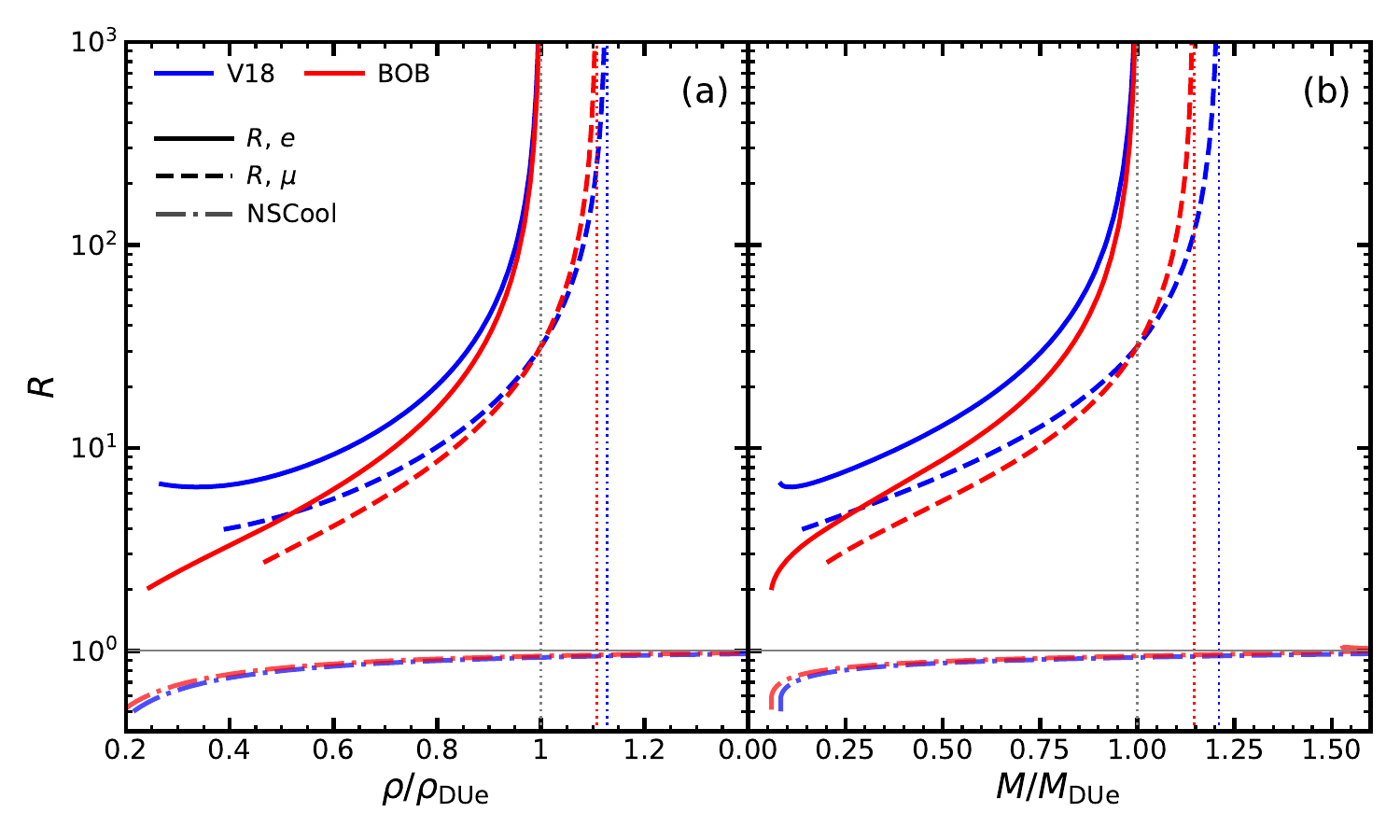}
}
\vspace{-6mm}
\caption{
MU rate modification factor $R$,
Eq.~(\ref{e:sbh}),
for the electron (solid) and muon (dashed) channels,
as a function of the normalized nucleon density (a)
or the associated NS mass (b)
for V18 and BOB EOSs.
The dotted vertical lines mark the DU$e$ and DU$\mu$ thresholds.
The dash-dotted curve shows the standard NSCool result $\alpha_n \beta_n$,
Eq.~(\ref{e:mua}).
}
\label{f:mmu}
\end{figure*}

\begin{figure*}[t]
\vspace{-12mm}
\centerline{\includegraphics[width=1.05\textwidth]{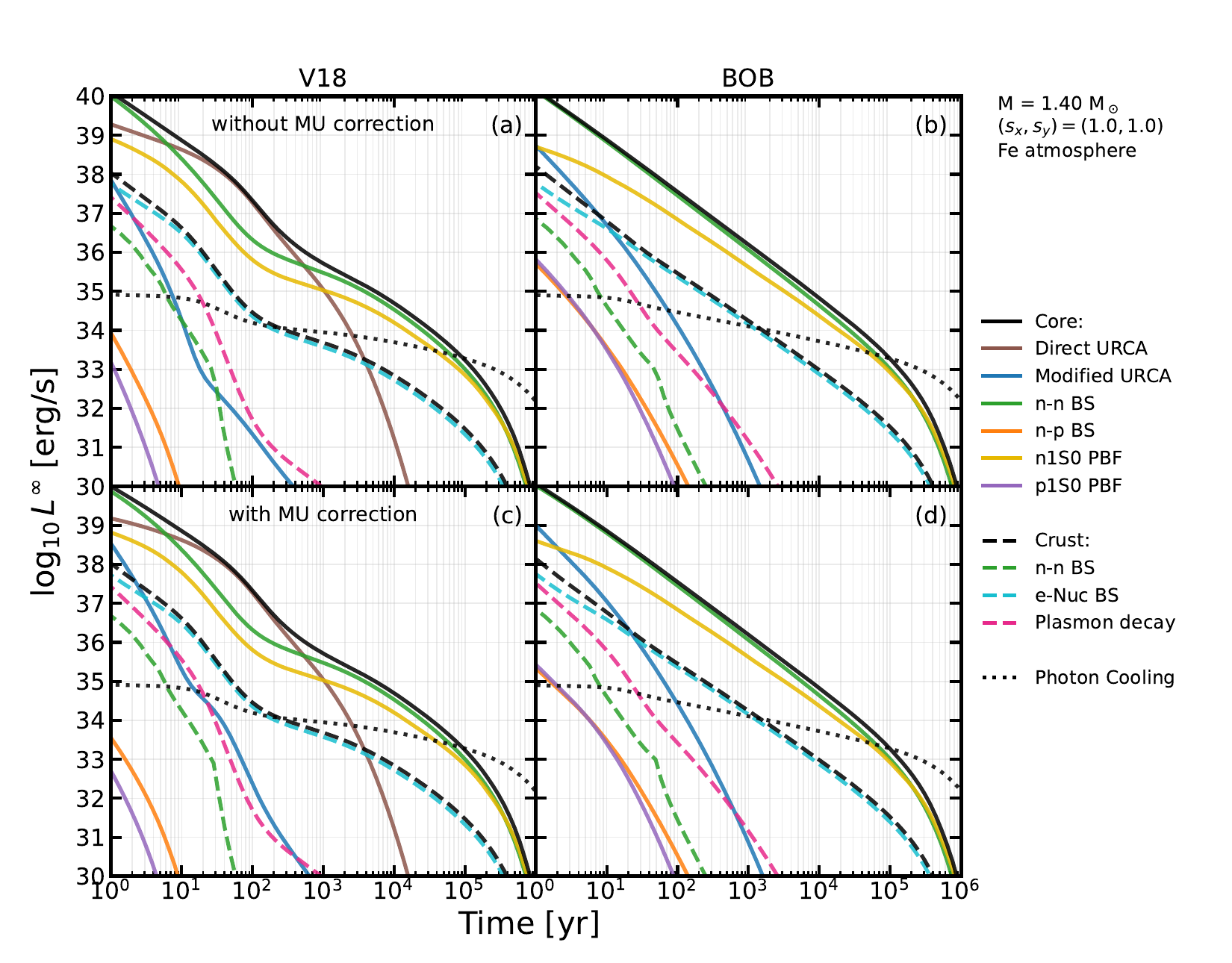}}
\vspace{-7mm}
\caption{
Detailed contributions of the various cooling processes to the total luminosity
for $M=1.4\ms$, $(s_x,s_y)=(1,1)$, and Fe atmosphere,
with the V18 (a,c) and BOB (b,d) EOS
without (a,b) and with (c,d) modification of the MU rate.
Shown are the luminosities of the various core (solid curves)
and crust (dashed curves) neutrino processes
and of the photon emission (dotted curves),
together with the total core and crust neutrino luminosities
(black solid and dashed curves).
The DU process is not operative for the BOB EOS at this mass.
}
\label{f:nupro}
\end{figure*}

\subsection{Gaps and mass distributions}
\label{s:cor}

At present, no quantitative information on the actual masses of the observed
cooling objects is available,
so a direct comparison between theoretically predicted and actual masses
in the cooling diagram is not feasible.
Under these circumstances,
one can {\em derive} a NS mass distribution
consistent with the outcome of a given cooling simulation
by counting the data points that fall within the interval
between two adjacent fixed-$M$ cooling curves
\cite{Wei19,Wei20,Das24}.
The resulting mass histogram can then be compared with NS mass distributions
obtained through different and independent theoretical approaches
\cite{Zhang11,Ozel12,Kiziltan13,Antoniadis16,Alsing18,Rocha19,Landry21}.

This procedure rests on two assumptions:
first, that the mass distribution of isolated NSs in the cooling diagram
does not differ from that of NSs in binary systems
\cite{Zhang11,Antoniadis16,Alsing18,Rocha19,Farrow19}
or of the entire NS population in the Universe;
and second, that the detection of
these sources is independent of their brightness
(Malmquist bias \cite{Wall12}).
Both assumptions are unlikely to be strictly satisfied,
as discussed in \cite{Wei20,Das24}.
A further fundamental limitation is the lack
of information on the atmospheres of the observed objects,
which necessitates additional theoretical assumptions in the analysis,
the simplest being the adoption of a fixed-atmosphere model.
Pending improved observational information on the data sources,
we nevertheless adopt this approach to derive the mass distribution.

The masses of the 53 cooling data points
(without UL data)
are assumed to be those predicted theoretically from their positions
in the cooling diagrams relative to the theoretical curves
shown in Fig.~\ref{f:cool}.
Figure~\ref{f:NSdist} displays the resulting mass histograms for different
choices of the p1S0 pairing parameters ($s_x,s_y$),
assuming a common Fe atmosphere
and comparing the results obtained with the V18 and BOB EOSs.

One observes that increasing $s_x$ or to a lesser degree $s_y$
shifts the centroid of the derived mass distributions to higher values,
since the cooling curves move upwards (warmer stars)
due to the increased suppression of the DU process.
Theoretical investigations of the NS mass distribution
\cite{Zhang11,Ozel12,Kiziltan13,Antoniadis16,Alsing18,Rocha19}
predict typically a median close to the canonical NS mass $1.4\ms$.
This is illustrated in the $(0.8,0.4)$ panel of Fig.~\ref{f:NSdist},
where we superimpose the theoretical results of
\cite{Zhang11,Rocha19}.
It can be seen that there is good agreement with the predictions of the V18 EOS,
in fact the values $(0.8,0.4)$
with corresponding $\m1s0=1.73\ms$ and $\Delta_\text{max}=1.2\;$MeV
are optimized for this purpose.

On the contrary,
with the BOB EOS no satisfactory agreement with the theoretical curves
can be achieved with any values ($s_x,s_y$).
The reason is the too large DU onset mass $\mdu=1.54\ms$ in this case,
which causes an abrupt onset of fast cooling at this mass,
and therefore a much larger spacing of the individual cooling curves
for larger masses than for smaller masses.
This clashes with the more or less smooth distribution of the data points.
This is a generic problem for any EOS with a DU onset close to $1.4\ms$:
the predicted mass histograms are strongly asymmetric around this value,
in contrast with both current cooling data and population models.
In our present nucleonic setup,
the only permitted DU onset masses have to lie far below $1.4\ms$,
as in fact is the case for the V18 EOS with $\mdu=1.10\ms$.

This important conclusion is also confirmed by the quantitative analysis
\cite{Wei20,Das24}
in terms of the rms deviations between the histograms
$\{N_i^\text{dat}\}$ in Fig.~\ref{f:NSdist}
and the expected
$\{N_i^\text{theo}\}$
for various theoretical distributions,
\be
 \delta N \equiv \sqrt{ \frac{1}{N^\text{dat}}
 \sum_i \Big(N_i^\text{dat} - N_i^\text{theo}\Big)^2 }  \:,
\label{e:rms}
\ee
where $i$ labels the mass bins and
$N^\text{dat}$ is the total number of data points contained
in the proper histogram.
The results are visualized in the heatmap shown in Fig.~\ref{f:heatmap}
for the various combinations $(s_x,s_y)$,
where we also indicate the optimal values $(s_x,s_y)$
(those predicting the smallest $\delta N$)
for each theoretical mass distribution.
Similar to Refs.~\cite{Wei20,Das24},
for a Fe atmosphere and the V18 EOS
the best agreement with most considered distributions
is obtained with $s_x\approx0.8$ and $s_y\approx0.4$,
which would also be consistent with microscopic investigations
of the 1S0 pairing gap \cite{Burgio21}, as discussed before.
On the contrary, for the BOB EOS no $(s_x,s_y)$ combination
reaches a comparably small value of $\delta N$,
mainly because its higher DU onset mass $\mdu=1.54\ms$
limits the fast-cooling component
that is required by the data
and is efficiently provided by the V18 EOS.

We close this section by showing in Fig.~\ref{f:best}
the best-fit,
$(s_x,s_y)=(0.8,0.4)$,
theoretical results,
where now the colors indicate the masses theoretically assigned
to the data points according to our analysis.

\subsection{Medium modifications}
\label{s:med}

We now quantify the impact of the in-medium enhanced MU rate
on the cooling evolution and on the mass-distribution analysis presented above.
For that purpose we have implemented in the {NSCool} code
the enhancement factor of \cite{Shternin18}, Eq.~(\ref{e:sbh}),
by replacing in the emissivity of the neutron branch,
Eq.~(\ref{e:mun}),
the standard matrix-element factor
$\alpha_n\beta_n=0.68\,[1.76-0.63(\rho_0/\rho_n)^{2/3}]$
by the density-dependent combination
$R=0.6R_A+0.2R_B$,
evaluated separately for the electron and the muon channel.
The modified expression is applied for $\rho\geq0.1\fm3$
and below the respective DU threshold,
where the parametrization is valid,
whereas outside this domain the standard rate is retained
in order to avoid the unphysical singularity at the DU onset.
The proton branch and the standard superfluid suppression factors
are not modified,
as suggested in \cite{Shternin18}.

The enhancement factor $R$, Eq.~(\ref{e:sbh}),
is displayed in Fig.~\ref{f:mmu}
for the electron (solid) and muon (dashed) channels
and for the V18 and BOB EOS.
Results are plotted as a function of the
normalized density $\rho\!/\rho_{\text{DU}e}$ (a)
and the associated NS mass (b).
The enhancement is already a factor of a few at half the threshold density
and grows unrestrained close to the DU onset,
which provides a smooth transition to DU cooling,
as advocated in \cite{Shternin18}.
The muon branch sets in only at
$\rho\approx0.4\rdu$,
but extends to higher density,
since its DU$\mu$ threshold lies slightly higher,
as discussed in Sec.~\ref{s:du}.
Plotted as functions of normalized units,
the results for the two EOSs nearly coincide:
the enhancement is essentially controlled by the proximity to the DU threshold,
and the strong EOS dependence seen in absolute density units
simply reflects the different values of $\rdu$ and $\mdu$ for both EOSs.
The dash-dotted curve shows the standard result $\alpha_n \beta_n$,
Eq.~(\ref{e:mua}),
implemented in NSCool,
which remains below unity and demonstrates that the enhancement
is dominated by the propagator effect.

The effect on the cooling evolution is illustrated in Fig.~\ref{f:nupro},
which decomposes the luminosity of a $M=1.4\ms$ star
with unscaled p1S0 gap, $(s_x,s_y)=(1,1)$, and Fe atmosphere
into the contributions of the various processes,
without (a,b) and with (c,d) the MU modification,
for the V18 (a,c) and BOB (b,d) EOSs.
For the V18 EOS the DU process is active ($M=1.4\ms>\mdu=1.10\ms$),
but is efficiently quenched by the p1S0 gap:
the $nn$ BS process, which involves no protons
and is therefore not affected by the proton superfluidity,
is the leading neutrino source at early and late times,
whereas DU is the largest contribution only around $t\sim10$--$100\,$yr;
the n1S0 PBF process is a comparable contribution throughout,
whereas the MU process is subdominant at all times,
suppressed by two order of magnitude relative to $nn$~BS initially
and much stronger later.
After $t\approx10^3\,$yr the $nn$ BS process dominates again
until the onset of the photon era at $t\approx10^5\,$yr.

For the BOB EOS the DU process is not operative at this mass
($\mdu=1.54\ms$),
and the $nn$ BS dominates the core luminosity at essentially all times,
exceeding the MU rate by about three orders of magnitude
at $t\approx10^2\,$yr and much more later,
because the p1S0 pairing suppresses the MU rate exponentially.
The n1S0 PBF process is the second contribution,
about one order of magnitude below the $nn$ BS rate.

Switching on the MU modification enhances the MU luminosity
of a $1.4\ms$ star by a factor of at most about twenty (V18)
or two (BOB) during the neutrino-cooling era,
but the surface photon luminosity changes by less than about 2\%
for $t\lesssim10^5\,$yr and by less than 10\% at later times,
which is far inside the observational uncertainties
of the cooling data.
The same holds for other masses
(for example $M=1.2\ms$ for the BOB EOS,
where the effect is even smaller).

We therefore conclude that in the currently favored scenario
-- fast DU cooling quenched by p1S0 pairing without n3P2 pairing--
medium modifications of the MU rates are practically insignificant
for the cooling evolution:
an enhanced MU rate could become visible only in NSs
without active DU cooling,
but precisely there the proton superfluidity quenches the MU process
and leaves the $nn$ BS as the dominant neutrino source
and the n1S0 PBF process as the second.
The cooling diagrams of Fig.~\ref{f:cool}
and the deduced mass distributions
of Figs.~\ref{f:NSdist} and \ref{f:heatmap}
are thus not modified,
and the main theoretical uncertainty
of the slow-cooling component of our models
resides in the $nn$ BS rate rather than in the MU rates.

\section{Conclusions}
\label{s:end}

We have presented
a global combined analysis of NS cooling
and mass distributions,
employing the most recent set of cooling data.
We confirm that those data demand a fast cooling mechanism like the DU process
(or alternative scenarios
\cite{Grigorian05,Blaschke07,Blaschke13,Shternin18,Suleiman23}
not considered in this work),
quenched by p1S0 pairing,
but no n3P2 pairing.

Comparing the two representative microscopic BHF EOSs employed here
(Argonne V18 and Bonn~B $NN$ potential
with compatible microscopic three-body forces),
the deduced NS mass distributions clearly favor the V18 EOS
with its low DU onset mass $\mdu=1.10\ms$,
combined with slightly reduced p1S0 pairing gaps,
$(s_x,s_y)\approx(0.8,0.4)$.
The BOB EOS with $\mdu=1.54\ms$
cannot reproduce equally well the considered theoretical mass distributions
for any gap combination,
because its fast-cooling component sets in at too high masses.
The confrontation with the cooling data therefore constrains
simultaneously the pairing gaps and the EOS,
in particular the onset density of the DU process,
which cannot be in the vicinity of the canonical mass $1.4\ms$.

We have also investigated the influence of the recently proposed
in-medium enhancement of the MU rates \cite{Shternin18}
on the cooling evolution.
Although the formal enhancement factor reaches very large values
close to the DU threshold,
its effect on the cooling curves and on the deduced mass distributions
turns out to be negligible:
once the DU process is quenched by the p1S0 superfluidity,
the same pairing gap suppresses the MU rates exponentially,
and the $nn$ BS remains the dominant neutrino source
in slowly cooling stars.
The main uncertainty of the slow-cooling component
of our NS cooling models therefore resides in the $nn$ BS rate
rather than in the MU rates.

Currently this method is mainly hampered by missing information
on masses and atmospheres of the cooling data,
which will constitute very effective constraints
when more abundant and precise data become available in the future.
This will also allow to pin down
the density range (onset density) of (blocked) DU cooling in the EOS,
which is another important degree of freedom
that was not studied in detail in this work.

Our general framework was a purely nucleonic scenario,
and exotic types of matter like hyperons or quark matter
were not considered in this work.
Those still present a formidable challenge to cooling calculations
due to their largely unknown microphysics ingredients relevant for cooling,
like cooling rates, heat capacities, transport properties,
and most of all superfluid properties.

\acknowledgments{
This work was partially funded by the
National Key R\&D Program of China No.~2022YFA1602303 and the
National Natural Science Foundation of China under
Grants Nos.~12205260, 12147101, and 11975077.
}

\newcommand{\araa}{Annu. Rev. Astron. Astrophys.}
\newcommand{\aap}{Astron. Astrophys.}
\newcommand{\apjl}{Astrophys. J. Lett.}
\newcommand{\epja}{EPJA}
\def\jcap{Journal of Cosmology and Astroparticle Physics}
\def\jcap{JCAP}
\def\jpg{J. Phys. G}
\newcommand{\mnras}{Mon. Not. R. Astron. Soc.}
\def\npa{Nucl. Phys. A}
\newcommand{\physrep}{Phys. Rep.}
\newcommand{\plb}{Phys. Lett. B}
\def\ppnp{Prog. Part. Nucl. Phys.}
\newcommand{\ssr}{Space Science Reviews}

\bibliography{coolmm} 

\end{document}